\def\kms{\;{\rm kms}^{-1} }

\def\kms{\;{\rm km}{\rm s}^{-1}  }

\def\ra{r_{\rm a}}

\def\rt{r_{\rm t}}

\def\sigmaP{\sigma_{\rm P}}

\def\spose#1{\hbox to 0pt{#1\hss}}
\def\lta{\mathrel{\spose{\lower 3pt\hbox{$\sim$}}
    \raise 2.0pt\hbox{$<$}}}
\def\gta{\mathrel{\spose{\lower 3pt\hbox{$\sim$}}
    \raise 2.0pt\hbox{$>$}}}
\documentclass{emulateapj}

\usepackage{amsmath}
\usepackage{amssymb}
\usepackage{graphicx}

\received{2004 May 24}
\begin{document}

\title{Kinematically Cold Populations at Large Radii in 
the Draco and Ursa Minor Dwarf Spheroidals}
\author{Mark I. Wilkinson\altaffilmark{1}, Jan T. Kleyna\altaffilmark{2}, N. Wyn Evans\altaffilmark{1}, Gerard F. Gilmore\altaffilmark{1}, Michael J. Irwin\altaffilmark{1}, Eva K. Grebel\altaffilmark{3}}
\altaffiltext{1}{Institute of Astronomy, Madingley Road,Cambridge, CB3 OHA, UK; markw@ast.cam.ac.uk}
\altaffiltext{2}{Institute for Astronomy, University of Hawaii, 2680 Woodlawn Drive, Honolulu, HI 96822}
\altaffiltext{3}{Astronomisches Institut, Universit\"{a}t Basel, Venusstrasse 7, CH 4102 Binningen, Switzerland}

\begin{abstract} 
We present projected velocity dispersion profiles for the Draco and
Ursa Minor (UMi) dwarf spheroidal galaxies based on 207 and 162
discrete stellar velocities, respectively. Both profiles show a sharp
decline in the velocity dispersion outside $\sim 30^\prime$ (Draco)
and $\sim 40^\prime$ (UMi). New, deep photometry of Draco reveals a
break in the light profile at $\sim 25^\prime$.  These data imply the
existence of a kinematically cold population in the outer parts of
both galaxies. Possible explanations of both the photometric and
kinematic data in terms of both equilibrium and non-equilibrium models
are discussed in detail. We conclude that these data challenge the
picture of dSphs as simple, isolated stellar systems.
\end{abstract}

\keywords{dark matter---galaxies: individual (Draco dSph, Ursa Minor
dSph)---galaxies: kinematics and dynamics---Local Group---stellar
dynamics}

\section{INTRODUCTION}

The dark matter dominated Local Group dwarf spheroidals (dSphs) have
emerged as valuable laboratories in which to test dark matter models.
Recently, the projected velocity dispersion profiles of the Fornax and
Draco dSphs have been obtained~\citep{mateo97,kleyna01}. Detailed
modelling of the discrete stellar velocities in Draco enabled the
hypothesis that mass follows light to be discarded at the $2.5\sigma$
level and suggested that the halo density $\rho(r)$ falls off more
slowly than the light distribution with $\rho(r) \sim
r^{-1.7}$~\citep{kleyna02}. In this {\it Letter}, we present new
observations of the Draco and Ursa Minor (UMi) dSphs which yield the
velocity dispersion profiles of both galaxies to the edge of their
light distributions.  The new data suggest the existence of {\it
kinematically cold populations} in the outer parts of both
galaxies. These data make it possible to test the validity of
isolated, equilibrium models of dSphs.

\section{OBSERVATIONS}

\subsection{Discrete Velocities}

We observed the Draco and UMi dSphs with the multifibre instrument
AF2/WYFFOS on the {\it William Herschel Telescope} on La Palma on
20-23 June 2003 and 6-11 May 2003, respectively. We drew our Draco
targets from the Sloan Digital Sky Survey, while we took our UMi
targets from our own KPNO 4m MOSAIC imaging. In each case, we
identified potential targets by drawing a polygon around the giant
branch of a $V, V-I$ color-magnitude diagram with a faint magnitude
limit of $ V < 20$.  The data were reduced using the WYFFOS-specific
{\tt WYFRED} data reduction package in {\tt IRAF} and cross-correlated
with the two blue-most lines of the Calcium triplet, in the same
manner as described by~\cite{kleyna02}. To determine the final member
list for each dSph we assumed that their velocity distribution was
Gaussian with a dispersion (including measurement error) of $\lta 12.5
\kms$.

Our final Draco dataset contains 112 velocities within $39 \kms$ of
Draco's mean velocity with a median velocity error of $2.4 \kms$.
The union of these data with the dataset of~\cite{kleyna01} contains
207 unique members with good velocities.  The median velocity of the
combined dataset is $-290.7^{+1.2}_{-0.6}
\kms$, where the uncertainties are obtained through bootstrap
re-sampling.  Our final UMi dataset has 144 stars within $36 \kms$ of
the mean velocity, with a median velocity error of $2.9 \kms$. Adding
the earlier UMi velocities of~\cite{kleyna03} produces a dataset with
162 member stars. The median velocity of the combined dataset is
$-245.2^{+1.0}_{-0.6} \kms$.

\begin{figure}[t]
\includegraphics[height=8.5cm]{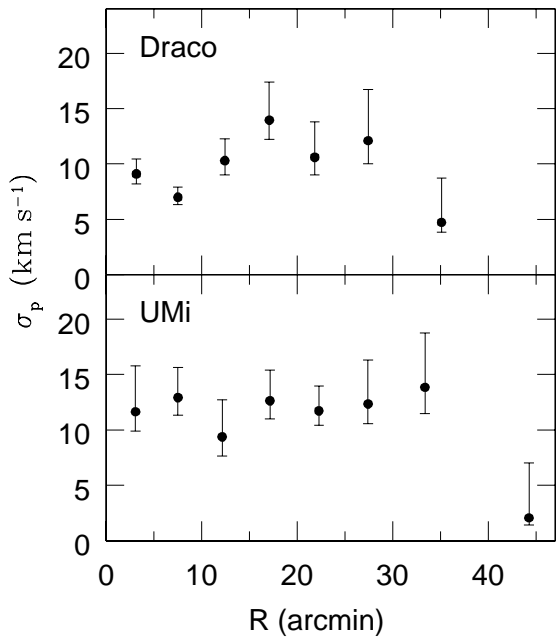}
\caption{Line of sight velocity dispersion profiles (with 1-$\sigma$ error
bars) for Draco and Ursa Minor. See text for a detailed discussion.}
\label{fig:velocities}
\end{figure}

For both Draco and UMi, at large radii the individual velocities of
the stars relative to the mean tend to decline with increasing
distance from the center. In the case of UMi, this is particularly
striking with six of the seven outermost stars lying within $1
\sigma$ of the mean bulk velocity of the dSph. Figure~\ref{fig:velocities} 
shows the radial variation of the the line of sight velocity
dispersion $\sigmaP$ in Draco and UMi. The projected dispersion drops
sharply at large radii. In each dSph, we detect no rotation beyond
that identified in previous work. Since our selection criterion for
dSph membership is based on the assumption of a Gaussian velocity
distribution, we would expect to discard less than one genuine member
in Draco (UMi) by imposing a $39 \kms$ ($36
\kms$) cut-off with the possible exception of extreme binaries. The
presence of some outliers just inside the velocity cut-off at large
radii suggests that we may, in fact, be retaining some non-members in
our sample. If 3 percent of our data belong to the Galactic foreground
(assumed to have a flat velocity distribution between our velocity
limits), then the overall dispersion of our sample would be increased
by about $1 \kms$. Thus, non-member contamination at large radii would
tend to strengthen our conclusion of a falling velocity dispersion. It
could be invalidated only if we had erroneously  removed large
velocity, bound stars from the sample.

\subsection{Photometry}
\label{sec:photom}
\begin{figure}[t]
\includegraphics[height=5cm]{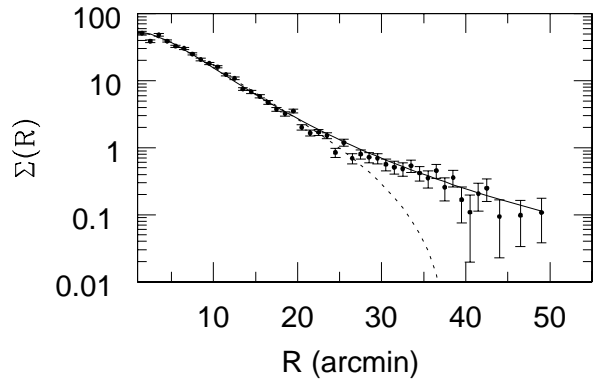}
\caption{Azimuthally-averaged surface brightness profile of Draco.
The solid and broken curves show, respectively, the best-fitting
Plummer profile and a~\cite{King62} profile fit to the data within
$25^\prime$.}
\label{fig:mike}
\end{figure}

Figure~\ref{fig:mike} shows the azimuthally averaged surface
brightness profile of Draco based on deep imaging to V$\approx$25 and
i$^\prime\approx$24 with the {\it Isaac Newton Telescope}. This has been
corrected for the effects of variable extinction using the reddening
map of~\cite{schlegel}. The limiting magnitude of our observations is
about two magnitudes fainter than that of the Sloan Digital Sky Survey
data used by~\cite{odenkirchen01} to determine the light profile of
Draco and permits more robust background subtraction in the outer
regions. In contrast to the ~\citeauthor {odenkirchen01} photometry,
the profile in Figure~\ref{fig:mike} shows a clear break at $\sim
25^\prime$. We note that a similar result has recently been claimed
by~\cite{kuhn}. The light profile of UMi displays a similar
feature~\citep[e.g.][]{IrHatz95,palma03} at about $34^\prime$. This
militates against the idea that the dSphs currently possess extended
halos~\citep[e.g. ][]{Stoehr02}.

Using simulations, \citeauthor{kjohnston} (\citeyear{kjohnston},
hereinafter J99) suggested that stellar systems in the Milky Way halo
will show breaks in their surface density profiles, beyond which
unbound or extra-tidal stars begin to predominate over bound
stars. Their results are superficially similar to our photometric data
on Draco and UMi, but there are some important differences. First, J99
found an enhanced velocity dispersion due to the extra-tidal stars and
second they argued for a slow fall-off ($\propto R^{-1}$) in surface
brightness beyond the break. \cite{kjohnston02} considered
non-circular dSph orbits and found a wide variety of profile shapes
and outer fall-off rates, with milder breaks and steeper fall-offs
occurring around the apocenters of more eccentric orbits. They also
found that the ratio of the break radius to the actual tidal radius
varies with orbital phase and eccentricity and is significantly below
unity near apocenter.

Also shown in Figure~\ref{fig:mike} are the best fitting~\cite{King62}
and Plummer profiles for the surface brightness profile of Draco. The
former fits the inner parts well, but requires an additional
extra-tidal population at $R > 25^\prime$ to mimic the break. The
latter provides a reasonable description of the entire profile. In the
rest of the {\it Letter}, we use these models to try to understand the
surprising data on the velocity dispersions of Draco and UMi.

\section{MODELLING}
\label{sec:mass_est}

\subsection{Equilibrium Models}
Using the simplifying assumptions of virial equilibrium and spherical
symmetry, the observable line of sight velocity dispersion
$\sigmaP$ as a function of projected radius $R$ is
\begin{eqnarray}
\sigmaP^2(R) & = & \frac{2}{I(R)} \int_R^\infty dr\, \nu(r) f(r) \frac{GM(r)}{r} \nonumber \\ 
& \times & \int_R^r dw\, \frac{w}{f(w) \sqrt{w^2-R^2}}
\Bigl[ 1- \beta(w) \frac{R^2}{w^2}\Bigr]
\label{eq:fundamental}
\end{eqnarray}
where $I(R)$ is the surface brightness and $\nu(r)$ is the stellar
luminosity density. This expression also involves the mass profile
$M(r)$ of the dark matter halo, and the stellar velocity anisotropy
parameter $\beta(r) = 1 - \langle v^2_\theta \rangle/ \langle v_r^2
\rangle$. The function $f(r)$ is the integrating factor for 
the spherical Jeans equation, namely $\exp[-\int_0^r dr\, 2
\beta(r)/r]$.

Under the assumption of isotropy ($\beta =0$),
eqn~(\ref{eq:fundamental}) becomes an Abel integral equation, which
can be inverted to give the mass profile $M(r)$ of the dark matter
halo~\citep[][ \S 4.2]{BT87}. Using an analytic fit to Draco's
projected dispersion together with either a Plummer or a King profile
in the Abel inversion, we find that the cumulative mass $M(r)$ becomes
unphysical ($dM/dr < 0$) beyond $r \sim 30^\prime$ . An isotropic
model with a Plummer or King profile cannot reproduce the observed
sharp decline in $\sigma_{\rm P}$ for Draco.  Analogous fits to the
dispersion profile and luminosity density of UMi lead to a similar
conclusion for its mass profile at the radius where its velocity
dispersion falls.

\subsubsection{A Sharp Change in the Velocity Anisotropy?}
\begin{figure}[t]
\includegraphics[height=7.5cm]{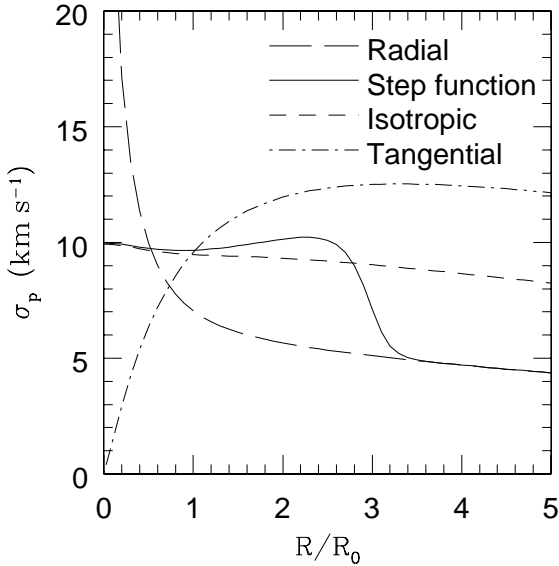}
\caption{Projected velocity dispersion for a Plummer light profile
(scale length $R_0$) in a dark matter halo assuming 4 different
velocity anisotropy laws.  Even an extremely sharp change in the
velocity anisotropy does not reproduce the decrease in anisotropy seen
in UMi.}
\label{fig:wyn}
\end{figure}
One possibility is that the velocity anisotropy changes abruptly from
isotropy $(\beta =0)$ in the inner parts to strong radial anisotropy
($\beta \rightarrow 1$) in the outer parts. This could cause a sharp
drop in the projected dispersion, even if the stellar density and dark
matter profile vary slowly and smoothly. We have investigated this
option using eqn~(\ref{eq:fundamental}) together with an anisotropy
parameter $\beta(r) = r^n/(r^n + \ra^n)$, which tends to a Heaviside
function $H(\ra)$ about the anisotropy radius $\ra$ as $n
\rightarrow \infty$.
Figure~\ref{fig:wyn} shows the projected velocity dispersion $\sigmaP$
assuming a Plummer model for the dSph luminosity density. The halo is
a `parabolic velocity curve' model as advocated by~\cite{Stoehr04} as
an excellent fit to sub-halos in high resolution simulations of the
Milky Way halo.  The observable dispersion is computed under the
assumptions of isotropy, extreme tangential or radial anisotropy
together with the Heaviside step function anisotropy. A very sharp
change in anisotropy can cause $\sigmaP$ to fall at $R \sim
\ra$. In fact, for a stellar population
with luminosity density $\nu \propto r^{-n}$ in a dark halo with an
underlying rotation curve behaving like $V(r) \sim r^{-\gamma}$,
$\sigmaP$ falls by a factor of $1/\sqrt{n+2\gamma -2}$ after a sudden
change to radial anisotropy.

For Draco, a smooth Plummer profile provides a reasonable fit to the
stellar density. Given the large error bar on the outermost data point
of the velocity dispersion, a sudden change in the anisotropy at large
radii could plausibly explain the data. For UMi, however, the
outermost dispersion data point is better constrained and the
photometry is less well represented by a smooth profile. A sharp
change in the anisotropy cannot by itself produce the drop in
$\sigmaP$, unless the stellar density or the dark halo also changes at
the anisotropy radius $\ra$.

\subsubsection{A Sharp Edge in the Light Distribution?}
\label{sec:sharp_edge}

If a dSph has a sharp edge in its light distribution, then $\sigmaP$
would necessarily fall to zero at projected radii approaching that
edge. Let us suppose that as $r \rightarrow \rt$, the stellar density
behaves like $\rho \sim (\rt-r)^n$.  Then, by expanding
eqn~(\ref{eq:fundamental}), it can be shown that
\begin{equation}
\sigmaP \sim C (\rt-R)^{1/2} R^{-(1\!+\!\gamma)/2},   
\end{equation}
The anisotropy $\beta$ and the fall-off in the stellar density $n$
alter the constant $C$, but not the scaling with distance from the
edge $\rt$. The dark halo has been assumed to possess an underlying
rotation curve behaving like $V(r) \sim r^{-\gamma}$, so that the case
$\gamma = 1/2$ (Keplerian rotation) corresponds to truncation of the
dark halo. So, irrespective of whether the dark halo is extended or
truncated, the velocity dispersion of the dSph must always go to zero
like $(R-\rt)^{1/2}$, if the stellar distribution has a sharp edge.

For both Draco and UMi, a sharp edge to the stellar distribution seems
at first sight inconsistent with the photometry.  However, as the
over-plotted King profile in Figure~\ref{fig:mike} shows, we can
associate a tidally limited model with Draco provided that the excess
of stars at $R \gta 25^\prime$ is interpreted as an extra-tidal (and
possibly unbound) population. If so, it is surprising that the
extra-tidal stars are kinematically colder than those in the main body
of Draco. This situation admits two possible explanations. First, the
tails could be kinematically cold, either intrinsically or due to
projection effects. This is not out of the question, as discussed
later in \S~\ref{sec:tides}. Here, we merely note that Liouville's
theorem tells us that the phase space density is conserved in the
absence of mixing, so that a stretching of material to form a tidal
tail must be accompanied by a corresponding contraction in velocity
space. Second, given the small numbers involved, is it possible that
we have missed stars associated with the extra-tidal enhancement?
This seems unlikely, as the density in the tails is higher than that
of Draco at the radii sampled by our outer velocity bin. However, the
presence of age or metallicity gradients could potentially lead us to
sample preferentially one population in the outer parts -- our Draco
photometric data provide some evidence that the blue and red
horizontal branches have different spatial distributions~\citep[see
also][]{klessen03}. In the case of UMi, the narrowness of the red
giant branch makes it more difficult to miss a dominant tail
population in the outer parts -- the weak age gradients identified
by~\cite{carrera} do not manifest themselves in the RGB. One might
also worry that the geometry of our WYFFOS pointings in UMi might miss
the areas dominated by the extra-tidal population. In fact, our outer
pointings were aligned with the major axis of the stellar distribution
and completely sample the elliptical region with a semi-major axis of
$50$ arcmin.

\subsubsection{Two-population Models for dSphs?}

It is worth considering whether Draco and UMi might contain two
kinematic populations: a hot, inner ``bulge-like'' component
surrounded by either a ``disk'' or ``halo'' component. One objection
to this model is that populations with different spatial distributions
must have formed under different conditions and would therefore not be
expected to display such similar stellar populations. However, it is
possible that the weak age gradients seen in UMi and the differences
in the spatial distributions of Blue and Red Horizontal Branch stars
in Draco indicate the presence of more than one old
population~\citep[see, e.g.,][]{harbeck01}.

The low projected velocity dispersion in the outer parts of both Draco
and UMi would appear to be consistent with the vertical velocity
dispersion of a thin disk. However, in each case the absence of an
observed rotation signal across the face of the dSph implies that we
must be observing the disk approximately face-on. For the
line-of-sight component of rotation to be smaller than our velocity
errors, any disk would have to be aligned to within $6^\circ$ of
face-on. This is in contradiction with the observed flattening of UMi,
which suggests that any disk must be close to edge-on. A
pressure-supported ``halo'' population might be a more plausible
candidate for the extra-tidal population. In Draco, the isopleths of
the stellar density distribution suggest that the extra-tidal
population within about $50$ arcmin has the same symmetry as the main
body of the dSph. However, the observed kinematics require that both
populations be truncated in order to give projected dispersions for
each population that fall rapidly in the outer parts. For UMi, this
model suffers from the additional difficulty of the significant
flattening required for both populations. If this is produced by
anisotropy in the velocity distribution (rather than by the effects of
external tides) it is at the extreme end of the range normally seen
for flattened systems without rotation.

\subsection{Tidal Sculpting of Draco and UMi?}
\label{sec:tides}

A natural explanation for a break in the light distribution of a dSph
is to invoke the external tidal field produced by the Milky Way. This
perturbs the outer regions of all Galactic satellites leading to the
escape of stars and the formation of tidal tails (e.g. J99). However,
in the case of Draco and UMi there are two problems associated with
this simple scenario. First, for Draco our mass estimate interior to
the break radius in the light is $1\times 10^8$M$_\odot$ while for UMi
it is $2\times 10^8$M$_\odot$.  Flattening of the gravitational
potential and velocity anisotropy causes uncertainty in these masses
by a factor of $\sim 2$. Both dSphs seem sufficiently massive that,
given any reasonable, assumed Milky Way halo profile, their current
tidal radii lie outside the observed light distribution.  This
suggests that neither Draco nor UMi is currently experiencing tidal
disturbance of its stellar distribution. Second, tidal effects
typically result in the heating of stellar populations and therefore
the velocity dispersion of the dSphs is expected to rise in the region
which has been influenced by tides, in stark contrast to the observed
data.

The difficulties might be resolvable if the Galactic orbits of Draco
and UMi are significantly elongated with peri-Galactica smaller than
$20$kpc. For Draco, such an orbit is not inconsistent with the
observed space-motion due to the large uncertainties on the measured
proper motion. The proper motion of UMi is currently better
constrained~\citep{schweitzer} and appears to rule out a deeply
plunging orbit. However, a preliminary measurement of the proper
motion based on HST data with a four-year baseline is consistent with
an elongated orbit (Piatek, priv. comm.). During perigalactic
passages, the tidal radii of the dSphs are smaller than at their
present locations.  If a dSph passes close to the disk of the Milky
Way, then the rapidly time-varying gravitational field disturbs its
outer parts. This generates a heated population of stars at the edge
of the dSph with an inflated velocity dispersion. If the orbit is
further constrained to have a perigalacticon passage after the mild
disk shock then the heated, extra-tidal population can drift away from
the dSph. Subsequently, as the dSph moves out to its current location,
the tidal radius increases again and the colder stars in the tidal
tails which have not had sufficient time to escape during the
pericenter passage are recaptured. This results in a population which
has the morphological appearance of a tidal tail but is kinematically
cold. It is also possible that the cold clump near the center of
UMi~\citep{kleyna03} is a projection of this cold, extra-tidal
population onto the face of the dSph.

This scenario places rather tight constraints on the properties of
both Draco and UMi as well as on their possible Galactocentric
orbits. First, their orbits must take them within $20$ kpc of the
Milky Way. Second, a simple estimate of their tidal
radii~\citep[following, e.g., ][]{kleyna01} shows that even at this
radius their tidal limits are large with respect to their stellar
populations unless their masses are below $5\times 10^7$M$_\odot$,
which is at the lower limit allowed by the data. Third, their orbits
must spend sufficient time near pericenter to allow the stars which
have been most heated during the close passage to escape before the
sphere of influence of the dSph engulfs them. We conclude that this
scenario is only plausible if Draco and UMi have somewhat lower masses
than previously estimated and have a low dark matter density outside
about $30$ arcmin. Further both dSphs must be on deeply plunging
orbits which take them close to the disk of the Milky Way. We are
currently performing $N$-body simulations to investigate this scenario
in more detail.

\acknowledgments 

\noindent  
MIW acknowledges support from PPARC. JTK is supported by the Beatrice
Watson Parrent\, Fellowship. The authors thank Michael Odenkirchen for
help with the Draco target list.


\end{document}